\documentclass[pra,preprint,showpacs,preprintnumbers,nofootinbib,sort&compress]{revtex4}

\usepackage{amsmath,amsfonts,amssymb,amsthm,graphicx}
\usepackage{bm}

\usepackage{xy}
    \xyoption{matrix}
    \xyoption{arrow}

\newcommand{\tmop}[1]{\operatorname{#1}}
\newcommand{\tmem}[1]{{\em #1\/}}

\newtheorem{theorem}{Theorem}
\newtheorem{conjecture}{Conjecture}

\newtheorem{lemma}{Lemma}
\newtheorem{corollary}{Corollary}

\theoremstyle{definition}

\theoremstyle{remark}

\def\til{\widetilde}

\newcommand{\cE}{\mathcal{ E}}

\newcommand{\cH}{\mathcal{ H}}

\newcommand{\cP}{\mathcal{ P}}

\newcommand{\cU}{\mathcal{ U}}

\def\R{\mathbb{R}}

\def\N{\mathbb{N}}

\def\bN{\mathbb{N}}

\def\g{\mathfrak{g}}

\newcommand{\fp}{{\mathfrak p}}

\newcommand{\su}{{\mathfrak{su}}}

\newcommand{\End}{\operatorname{End}}

\newcommand{\Mod}{\operatorname{Mod}}

\def\<{\langle}
\def\>{\rangle}

\def\til{\tilde}
\def\te{\text}

\def\f{\frac}

\newcommand{\lrp}[1]{\left( {#1} \right) }

\def\be{\begin{equation}}
\def\ee{\end{equation}}
\def\bea{\begin{eqnarray}}
\def\eea{\end{eqnarray}}
\def\beas{\begin{eqnarray*}}
\def\eeas{\end{eqnarray*}}

\newcommand{\er}[1]{\eqref{#1}}
\newcommand{\bel}[1]{\begin{equation}\label{#1}}

\begin{document}

    \title{On the number of representations providing noiseless subsystems}

    \author{William Gordon Ritter}
    \affiliation{Harvard University, Department of Physics}
    \date{July 24, 2005}

\begin{abstract}
This paper studies the combinatoric structure of the set of all
representations, up to equivalence, of a finite-dimensional
semisimple Lie algebra. This has intrinsic interest as a previously
unsolved problem in representation theory, and also has applications
to the understanding of quantum decoherence. We prove that for
Hilbert spaces of sufficiently high dimension, decoherence-free
subspaces exist for almost all representations of the error algebra.
For decoherence-free subsystems, we plot the function $f_d(n)$ which
is the fraction of all $d$-dimensional quantum systems which
preserve $n$ bits of information through DF subsystems, and note
that this function fits an inverse beta distribution. The
mathematical tools which arise include techniques from classical
number theory.
\end{abstract}


    \maketitle

\section{Introduction}

A variety of schemes for protecting quantum information have been
developed, including quantum error correction codes
\cite{Bennett,Gottesman,Knill:97a,Calderbank}, decoherence free
subspaces
\cite{Zanardi:97a,Zanardi:97c,Zanardi:98a,Lidar:PRL98,Lidar:1998nd},
noiseless subsystems  \cite{Knill:99a}, bang-bang decoupling
\cite{Viola:1998a}, and topological quantum computation
\cite{TOP}. The first four of these techniques are closely related
to each other and can be described in a simple unified framework
based on representations of the algebra of errors
\cite{Knill:99a,Zanardi:99d,RZ:2000}. More recently, it was shown
\cite{ZL} that topological quantum computation is also related to
the error-algebra framework.

Decoherence-free subspaces and subsystems have already been realized
in many important laboratory experiments, of which we mention a few.
Consider a system exhibiting electromagnetically induced
transparency \cite{Harris:97}, such as a tenuous vapor of 3-level
atoms in the ${\rm \Lambda} $ configuration (e.g. Strontium),
contained in an optical resonator, as described in
\cite{LidarWhaley}. Decoherence in this system arises from
spontaneous emission, which corresponds to a transition from the
excited state to any of the lower lying states. Since spontaneous
emission occurs only from the excited state, those states in the
orthogonal complement are a decoherence-free subspace.

Zanardi and Rasetti \cite{Zanardi:97c} showed that
decoherence-free subspaces exist in the phenomenologically
important \emph{spin-boson model} of $N$ spins coupled to a
bosonic quantum field, under the assumption that the coupling
constants are the same for several of the spins. DFSs also arise
as a consequence of a collective coupling to the environment in
the Jaynes-Cummings Hamiltonian for $N$ identical two-level atoms
coupled to a single mode radiation field
\cite{Duan:98b,Duan:98c,Zanardi:98a}.

In this short note, we consider all possible representations of the
error algebra $\su_3$, and determine how many of these admit
decoherence-free subspaces and noiseless subsystems of various
sizes. We prove that for quantum mechanical systems of sufficiently
high dimension, decoherence-free subspaces exist for almost all
representations of the error algebra. Therefore, our results provide
further arguments in favor of the error algebra framework.

Consider a quantum mechanical system with a finite-dimensional
Hilbert space $\cH$, and let $D = \dim (\cH)$. This class of systems
includes all spin systems, and the state space of any quantum
computer. Further suppose that our system is an {\tmem{open quantum
system}}, which means that it interacts with another system, called
the ``environment,'' and the latter has a Hilbert space $\cE$.
Unitary evolution for the combined system $\cH \otimes \cE$ implies
non-unitary evolution for the reduced density matrix $\rho :\cH
\rightarrow \cH$ obtained by tracing out the degrees of freedom
associated to $\cE$.

The interaction Hamiltonian $H :\cH \otimes \cE \rightarrow \cH
\otimes \cE$ which describes the system-environment coupling
necessarily takes the form
\begin{equation}
  H = \sum_{a = 1}^k F_a \otimes B_a \label{interaction} \, ,
\end{equation}
where $\{ F_a : a = 1 \ldots k \}$ is a set of Hermitian operators
on $\cH$.

If the commutators $[F_a, F_b]$ are expressible as linear
combinations of the $F_a$,
\bel{liealg}
  [ F_a, F_b ] = i \sum_c {f_{a b}}^c F_c,
\ee
then according to a classic result on decoherence
\cite{Lidar:PRL98}, a subspace $V \subset \cH$ is decoherence-free
if and only if $V$ decomposes as a direct sum of singlets under
the action of the $F_a$ operators, where a \emph{singlet} is
defined to be a 1-dimensional invariant subspace of $V$.

Assuming \er{liealg}, we may define an abstract Lie algebra $\g$
with generators $\{ x_a \}$, where $a = 1\ldots k$ and relations
$[x_a, x_b] = i {f_{ab}}^c x_c$. There is then a unique linear map
\bel{phiF}
    \phi_F : \g \to \End(\cH)
\ee
with $\phi_F(x_a) = F_a$. The map $\phi_F$ is a unitary
representation of the Lie algebra, since the $F_a$ are Hermitian
by construction. The reason we make a point to distinguish the
abstract Lie algebra $\g$ from its generators in a particular
representation is that we wish to consider the effect of changing
the representation without changing the symmetry algebra $\g$.

For some dimension-dependent number $p$ between 0 and 100,
decoherence-free subspaces exist in exactly $p$ percent of all
$D$-dimensional open quantum systems with errors acting in all
possible ways. In order to allow concrete calculations, we restrict
attention to the error algebra $\mathfrak{g} = \su_3$; the
calculations for other Lie algebras differ only in technical
aspects.

If $p < 5\%$, one might be tempted to question the utility of the
error-algebra framework, since representations which satisfy the
necessary criteria would seem to be sparse in the set of all
representations. On the other hand, if $p$ is close to 100\%, then
reliable methods for protecting quantum information from
$\su(3)$-generated errors are plentiful. Therefore, the precise
value of $p$ is of fundamental importance. In this paper, we show
that a reasonable answer for $D$ on the order of $10^1 \sim 10^2$ is
$p \approx $ 70--90 percent, and $p$ increases with $D$. Also, $p$
seems to asymptotically approach 100\%, so in higher dimensional
systems there are many more representations which protect quantum
information (provided that the error algebra does not also increase
in dimension).

Our results imply that in most dimensions, decoherence-free
subspaces exist for almost all representations of the error algebra.
However, there are certain dimensions to watch out for: in $D = 6$,
only three of eight total representations contain a singlet, or 37.5
percent. Therefore, it's especially difficult to preserve quantum
information in a 6-dimensional Hilbert space, when the errors are
generated by $\su_3$. Interestingly, this is related to the fact
that 3 divides 6.

While our results are mathematically rigorous, any connection of
these results to laboratory studies of decoherence must be
classified as highly speculative, for reasons which we now explain.
A finite-dimensional open quantum system is specified by a Hilbert
space $\cH$, system Hamiltonian $H_s \in \End(\cH)$, error algebra
$\g$, and a representation $\phi: \g \to \End(\cH)$. This paper
merely determines (for fixed $\cH, H_s, \g$) the number of
representations $\phi$ that support protection of quantum coherence,
and the statistical distribution of the sizes of noiseless
subsystems over the space of possible representations of the error
algebra. We do not discuss the probability that any particular
physical system evolves according to the dynamics specified by such
a representation.

As a thought experiment, consider an initially isolated $D$-level
quantum system, which is then brought into contact with an
environment in a controlled way, so that (we suppose) the error
algebra is known to be $\su_3$. One would like to know the true
likelihood of encountering a decoherence-free subspace in this
situation. The numbers reported in Fig.~\ref{Fig:Convergence} would
answer this question if, by some unknown mechanism, Nature selects
one of the available $\su_3$-representations at random, with a
\emph{uniform} weighting. As no mechanism which operates this way is
currently known, it seems impossible at present to connect
Fig.~\ref{Fig:Convergence} with experiment.

On the other hand, our results show where to look in order to find
examples with noiseless subsystems, and also highlight situations in
which DF subspaces or noiseless subsystems are rare. As such, these
results should help theorists to construct models in which quantum
coherence is protected.

\section{An exact formula for the number of su(3) representations in a fixed dimension }

\subsection{The Exact Formula}

Let $\xi (n)$ equal the number of irreducible representations of
the Lie algebra $\su(3)$ in dimension $n$. We compute an explicit
formula for this function in Appendix \ref{AppA}.

We now fix a dimension $D$ and determine certain facts about the
structure of all possible representations (including reducible
ones) in that dimension. In particular, we determine the fraction
which contain a singlet. Any representation in dimension $D$ may
be decomposed into irreducibles, and the dimensions $p_i$ of those
irreducible components determine a partition \cite{Andrews} of the
integer $D$. We say that the representation has {\tmem{shape}}
corresponding to a given partition if the dimensions of its
irreducible components are precisely the integers appearing in
that partition.

For a partition $\fp = ( p_1, \ldots, p_n )$ of $D$, a naive guess
for the number of representations with shape $\fp$ would be
$\prod_{i = 1}^n \xi ( p_i )$. This guess is correct if and only if
$\fp$ does not contain repetitions. For example, if $p_1 = p_2$ and
\mbox{$\xi ( p_1 ) > 1$} then we are over-counting. In order to
properly count these cases, we use a standard combinatorial
function, which we now explain.

If $p_i$ is repeated $n_i$ times in a given partition, we think of
the $k = \xi ( p_i )$ different $p_i$ dimensional irreps as
``letters'' in an alphabet. The set of distinct $n_i$-fold direct
sums of these irreps is in 1-1 correspondence with the set of
length $n_i$ words in $k$ letters, where order of letters in a
word is not important.

\begin{lemma}
  \label{lemma:P}Let $S ( n, k )$ equal the number of length $n$ strings from
  an alphabet of $k$ letters, with order not important. Then
  \[
        S( n, k ) = \binom{n + k - 1}{n} .
  \]
\end{lemma}

The recursion relation $S( n, k ) = \sum_{\ell = 0}^n S( n - \ell,
k - 1 )$ follows immediately, and gives an efficient way of
calculating the values of $S$. We now return to our objective of
counting the total number of $\su(3)$ modules in dimension $D$.

Let $\mathcal{P}( D )$ denote the set of all partitions of $D$.
Given a set $H$ of non-negative integers, let $\cP(H, n)$ denote
the set of partitions \mbox{$n = \sum_i p_i$} with $p_i \in H$ for
all $i$, so that $\cP(H,n) \subset \cP(n)$ for all $H \subset \N$.
Let $R_3$ be the set of possible dimensions of an $\su_3$-module,
\begin{align}\label{R3}
    R_3 &= \{ d \in \N :  \xi(d) \ne 0 \}
    \cr &=
    \{ 1, 3, 6, 8, 10, 15, 21, \ldots \} \, .
\end{align}
Therefore computation of $R_3$ reduces to computation of $\xi$,
which is done in the appendix.

\begin{theorem} \label{thm:master}
  Let $\Mod ( \su_3, D )$ denote the total number of $\su(3)$
  modules in dimension $D$. Assume that each partition $\fp$ of $D$ is to be
  expressed in the form $D = \sum_{i = 1}^n n_i p_i$  where the $p_i$ are all
  distinct. Then
\bel{eq:master}
  \Mod ( \su_3, D ) = \sum_{\fp\, \in\, \cP(R_3,\, D )}\,
  \prod_{i=1}^n \binom{n_i + \xi ( p_i ) - 1}{n_i}  .
\ee
\end{theorem}

When $\xi \neq 0$, its most likely value is $\xi = 2$, and then
the binomial coefficient simplifies to $( n + 1\ \tmop{choose}\ n
) = n + 1$. In that case, the partitions that contribute the most
are those which maximize the product $\prod_{i = 1}^n ( n_i + 1
)$, which is the same as maximizing $\prod n_i$. Therefore, the
largest terms in the sum (\ref{eq:master}) are those that do not
contain singlets; however, the terms which do contain singlets are
more numerous. The competition between these two types of terms
determines the fraction of representations which contain a
singlet, which we analyze in the next section.

\subsection{An algorithm for efficient computation}
\label{sec:efficient}

The sum in \er{eq:master} is over $\cP(R_3,\, D )$, the set of
partitions of $D$ with parts in $R_3$, a specific subset of the
positive integers. For efficient computation of \er{eq:master}, it
is essential to have an algorithm which lists only the partitions
we are interested in, without having to first list all partitions
and then filter out those which do not meet our criteria. As the
number of partitions grows exponentially according to the
Hardy-Ramanujan formula, any reduction in the number of terms is
crucial to make computation of $\Mod ( \su_3, n )$ even possible.

Fix a set of non-negative integers,
\[
    H = \{1, n_1, n_2, \ldots\}
    \subset \bN
\]
with $n_i < n_j$ if $i < j$ and $n_1 > 1$. We present a simple
algorithm for explicitly computing the set of partitions of $n$
whose parts lie in $H$.

A partition $n = \sum_{i=1}^m k_i$ is said to be in \emph{reverse
lexicographic order} if $k_1 \geq k_2 \geq \ldots \geq k_m$. We
will refer to a partition which is in reverse lexicographic order
as an \emph{ordered partition} for brevity.

Let $P(n,k)$ compute all ordered partitions of $n$ that begin with
a number between 1 and $k$. Let
\[
    P(n) = \bigcup_{k \geq 0} P(n,k)
\]
denote all ordered partitions. Let
\[
    P_H(n,k) = \{ (k_1, \ldots, k_m) \in P(n,k) \mid k_i \in H \
    \forall \, i \}
\]
denote the subset consisting of those ordered partitions whose
elements come from $H$. Similarly define $P_H(n)$.

If $\kappa = \{k_1, \ldots, k_p\}$ is an ordered partition of an
integer $k$, and $\ell > 0$ is a positive integer, let $\ell \vee
\kappa$ denote the ordered partition of $k+\ell$ given by
prepending $\ell$ to $\kappa$, i.e.
\[
    \ell \vee \kappa
    =
    \{\ell, k_1, \ldots, k_p\} .
\]
We extend this notation to sets of partitions in the obvious way,
so that $\ell \vee P(n)$ denotes $\{ \ell \vee \kappa : \kappa \in
P(n)\}$.

\begin{theorem}\label{fast} The set of partitions whose parts come
from $H$ is given by
\[
    P_H(n) = \bigcup_{k=2}^{n-1}
    \left\{ \begin{matrix} (n-k) \vee P_H(k,n-k) & \te{ if }\ n-k
    \in H \\
    \emptyset & \te{ otherwise} \end{matrix}
    \right. \ .
\]
\end{theorem}

Theorem \ref{fast} gives a fast recursive algorithm for
enumerating only the partitions of $n$ with parts in $H$. Even on
a fast computer, the naive algorithm of enumerating all partitions
and subsequently choosing those with the right properties will
lead to space problems due to the exponential growth of \er{hr}.

\subsection{Asymptotics for large $n$}

Formula (\ref{eq:master}) gives a way of evaluating $\Mod(\su_3,
n)$ numerically, although this is not efficient (or even possible
at all for large $n$) as it involves a sum over partitions. If
$p(n)$ denotes the number of all partitions for $n$, the
Hardy-Ramanujan asymptotic formula gives
\bel{hr}
  p (n) \sim \frac{1}{4 n \sqrt{3}}
  \exp\left[ \pi \sqrt{2n/3} \right]
\ee
Good references on this include \cite{HardyWright,Andrews}.

\begin{conjecture} \label{conj}
$\Mod ( \su_3, n )$ is given, asymptotically for large $n$, by a
formula similar to the Hardy-Ramanujan formula \er{hr}.
Specifically,
\[
    \Mod( \su_3, n ) \sim \frac{a}{n} \exp ( b n^c )
\]
for some positive, real constants $a, b, c$.
\end{conjecture}

The inspiration for this conjecture was Meinardus' theorem (see
\cite{Andrews} and references therein), which states roughly that
a large class of partition functions have exponential behavior
which generalizes that of the Hardy-Ramanujan function.

\begin{figure}
    \includegraphics[width=5in]{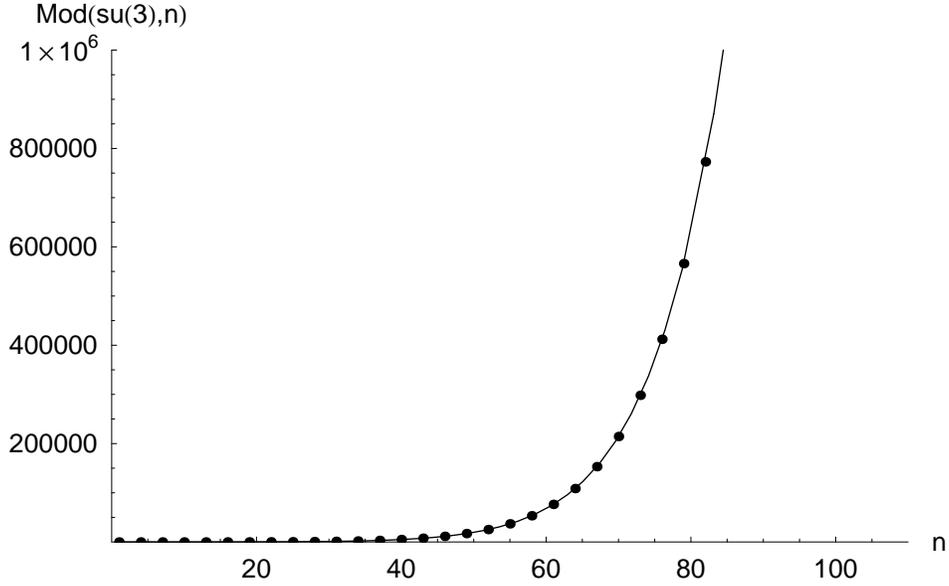}
    \centering
    \caption{\label{fig:Mod}Graph of \mbox{Mod($\su(3)$, $n$)} vs. $n$,
    with only points \mbox{$n \equiv 1$} \mbox{$(\te{mod}\ 3)$} shown, fitted by
    the curve $(a/n)\exp ( b n^c )$ with parameter values
    given by Eqn.~\er{params}. The points with $n \equiv 0$
    (mod 3) could also be fitted to a smooth curve, as could
    those congruent to 2 (mod 3); those curves have slightly
    different parameter values.}
\end{figure}

Nonlinear regression over the interval $[1,110]$ determined that
\bel{params}
    ( a, b, c )
    \sim
    (0.0771591, 2.70605, 0.459802)
\ee
which is surprisingly close to the values 0.14, 2.57, and 0.5
given by (\ref{hr}). Fig.~\ref{fig:Mod} shows the curve $(a/n)
\exp ( b n^c )$ with the approximate values \er{params}, together
with points representing the exact values of $\Mod( \su(3), n )$
computed using Theorems \ref{thm:master} and \ref{fast}, for $n
\equiv 1$ $(\te{mod}\ 3)$.

\section{Statistics of the set of all representations}

\subsection{Singlets} \label{sec:singlets}

As an example to illustrate formula (\ref{eq:master}), consider $D
= 6$. There are 11 partitions of 6, only 4 of which correspond to
direct sum decompositions of representations of $\su_3$:
\[
    \begin{array}{llll}
  p = \{ 6 \}               & & &     2\ \tmop{representations} \\
  p = \{ 3, 3 \}            & & &     3\ \tmop{representations} \\
  p = \{ 3, 1, 1, 1 \}      & & &     2\ \tmop{representations} \\
  p = \{ 1, \ldots, 1 \}    & & &     1\ \tmop{representation}
    \end{array}
\]
Therefore in $D = 6$, three of eight total representations contain
a singlet, or 37.5  \% . By contrast, for $D = 5$, because there
are no representations of $\su(3)$ in dimensions 2, 4 and 5, only
two partitions contribute. These are $5 = 3 + 1 + 1$ and $5 = 1 +
\cdots + 1$ giving 2 and 1 representations respectively. Thus for
$D = 5$, all three of the three available representations contain
a singlet.

\begin{figure}
    \includegraphics[width=5in]{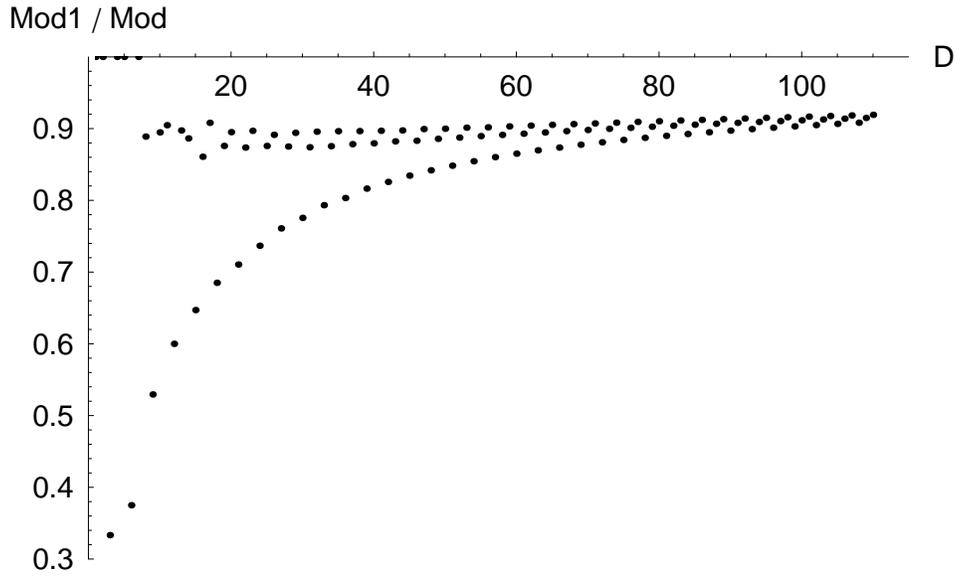}
    \centering
    \caption{\label{Fig:Convergence}Fractions of
    $\su(3)$ representations that contain a singlet vs. $D$, for $D \leq 110$.
    There are three distinct curves corresponding to the
    distinct congruence classes of $D$ modulo 3. }
\end{figure}

Define $\Mod_1 ( \su_3, D )$ by the same formula as
(\ref{eq:master}) but with the sum restricted to partitions
containing 1 as one of the elements; the fraction of $\su_3$
representations that contain a singlet is then ${\Mod_1} /
{\Mod}$. In Fig.~\ref{Fig:Convergence}, we plot ${\Mod_1} /
{\Mod}$ as a function of $D$. The creation of this plot could not
have been possible without the computational speedups suggested by
Theorems \ref{thm:master} and \ref{fast}, which is most likely why
it has never appeared in the literature before.

Fig.~\ref{Fig:Convergence} has many fascinating features, which
should be of interest to pure mathematicians as well as to
physicists. Beginning around $D = 10$, one can notice three series
in the plot which seem to converge. These three series correspond
respectively to the cases $D \equiv 0, 1, 2\ ( \te{mod}\ 3 )$.

Within each series, the points are very regular and seem to line
themselves up along a smooth curve, which at first seems
mysterious. It's possible that an explanation for this behavior
may be provided by some generalization of Conjecture \ref{conj}.
If we fix the residue class of $n$ modulo 3, and then find that
\[
    \Mod_1( \su_3, n ) \sim (a_1/n) \exp ( g(n) )
\]
for some function $g$, then we have
\[
    \Mod_1 / \Mod
    \sim
    \f{a}{a_1} \exp\lrp{b n^c - g(n)} .
\]
If $g(n)$ is such that $b n^c - g(n)$ is negative but approaches
zero monotonically from below, then we obtain the behavior
observed in Fig.~\ref{Fig:Convergence}.

\subsection{Noiseless Subsystems} \label{sec:subsystems}

Decoherence-free subspaces do not provide the most general method
for decoherence-free encoding of quantum information. Knill,
Laflamme, and Viola \cite{Knill:99a} discovered a method for
decoherence-free encoding using \emph{subsystems}. Zanardi soon
thereafter realized that this allowed a unification of many
seemingly unrelated ideas for reducing decoherence
\cite{Zanardi:99d}. Kempe \textit{et al.} developed a general
theory of universal quantum computation based on the
decoherence-free (or \emph{noiseless}) subsystem concept
\cite{Kempe:00}.

\begin{figure}
    \includegraphics[width=5in]{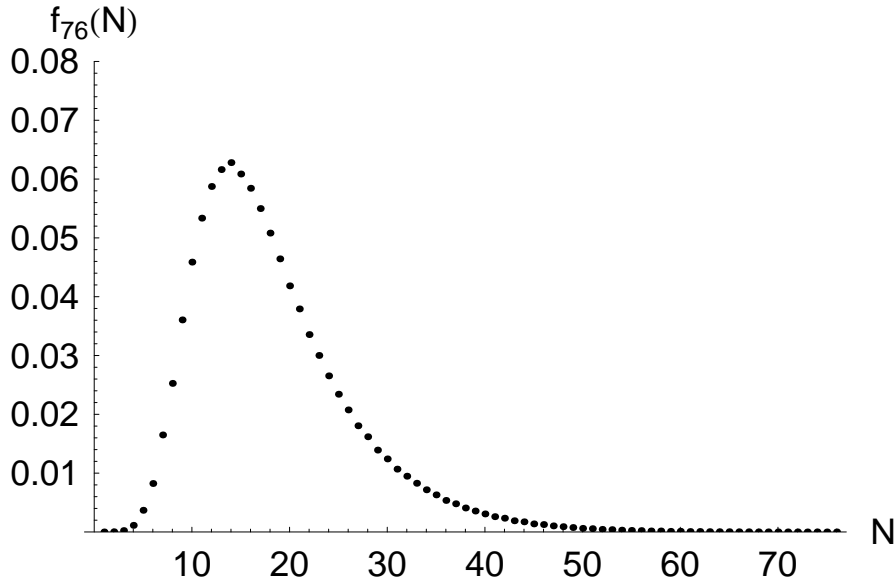}
    \centering
    \caption{\label{Fig:ss76}the fraction $f_d(N)$ of all
        representations in dimension $d=76$ which
        preserve an $N$-dimensional noiseless subsystem,
        as a function of $N$. The peak is at $\sim d/5$.}
\end{figure}

\begin{figure}
    \includegraphics[width=5in]{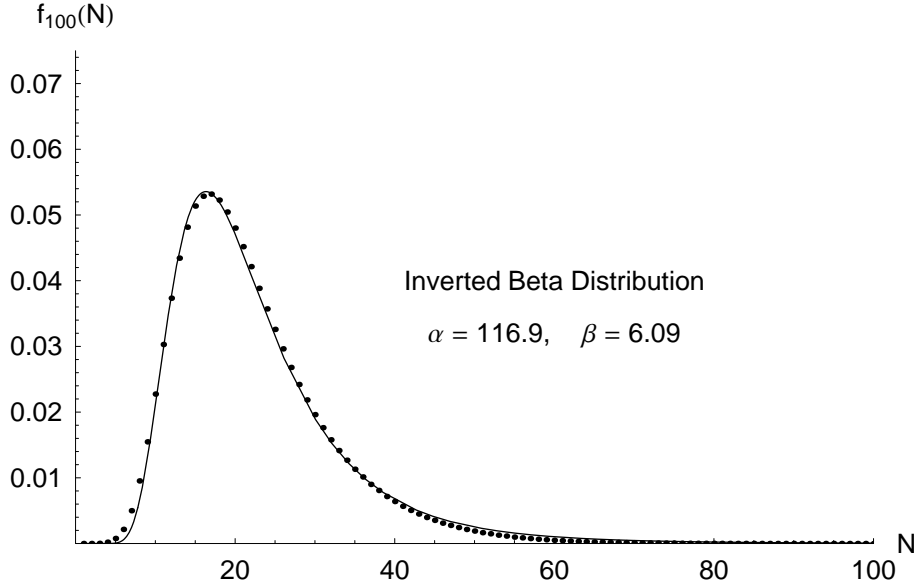}
    \centering
    \caption{\label{Fig:ss100}
        Data points represent exact computations of
        $f_{100}(N)$, the fraction of
        representations in dimension $d=100$ which
        preserve an $N$-dimensional noiseless subsystem.
        The points are fitted to an inverted beta
        distribution (see equation \er{eq:ibeta})
        with $\alpha \sim 116.907$ and $\beta \sim 6.091$.
        Comparing with figure \ref{Fig:ss76}, which has $d=76$,
        we note that in both cases, the peak is near $d/5$.}
\end{figure}

Irreducible representations of the error algebra do not possess
noiseless subsystems. We therefore devote this section to
determining how many representations contain noiseless subsystems
of various sizes. Our results are presented in
Figs.~\ref{Fig:ss76} and \ref{Fig:ss100} and the surrounding
discussion.

A representation $\phi$ of $\g$ lifts to a unique associative
algebra homomorphism $\tilde{\phi}$ of the universal enveloping
algebra $\cU(\g)$, by the universal property most elegantly
expressed in the commutative diagram
\bel{commdiag}
    \xymatrix{ \g \ar^i[r] \ar_{\phi}[dr] & \cU(\g) \ar^{\tilde{\phi}}[d] \\
    \ & \End(V) }
\ee
The action of $\tilde{\phi}$ is simply to convert the tensor
product to matrix multiplication, i.e. \mbox{$\tilde{\phi}(x
\otimes y)$} $=$ \mbox{$\phi(x) \cdot \phi(y)$}, etc.

\begin{lemma} \label{lemma:surjective}
If $\phi$ is an irreducible faithful representation and if $\g$ is
a semisimple Lie algebra, then $\tilde{\phi}$ is surjective.
\end{lemma}

(The proof of lemma \ref{lemma:surjective} is an easy exercise.)

Each irreducible representation $\phi$ will have the property that
$\til{\phi}(\cU(\g))$ will be the full matrix algebra $\End(V)$.
On the other hand, noiseless subsystems take advantage of a basis
in which $\til{\phi}(\cU(\g))$ is block diagonal. Therefore, these
blocks correspond to a decomposition of the Hilbert space $\cH$
into irreducible subrepresentations of the original representation
coming from the $F$-operators,
\[
    \cH = \Big(
        \bigoplus^{n_1} V_{d_1} \Big) \oplus \dots
        \oplus \Big( \bigoplus^{n_r} V_{d_r} \Big)
\]
where the subspaces are labelled by dimension, $\dim(V_{d_i}) =
d_i$. In this situation, we can send an $N$-dimensional vector,
where $N = \sum_i n_i$, through the quantum channel unaffected by
decoherence.

If the representation of the $F$-operators does indeed have such a
decomposition, then it corresponds naturally to a partition
\[
    D = \sum_{i=1}^r n_i d_i
\]
One may now compute the percent of all representations in
dimension $D$ which preserve an $N$-dimensional vector, for
various $N < D$. Recall that $R_3$, which was defined in \er{R3},
denotes the set of positive integers which are the dimension of an
irreducible $\su_3$ module.

\begin{theorem}
Let each partition $\fp$ of $D$ be expressed as $D=\sum_{i = 1}^n
n_i p_i$. The number of representations in dimension $D$ which
preserve an $N$-dimensional vector is given by
\[
    \sum_{  {\scriptstyle\fp\, \in\, \cP(R_3,\, D )}
            \atop
            {\scriptstyle\te{ with } \, N = \sum_i n_i}}
    \
    \prod_{i=1}^n \binom{n_i + \xi ( p_i ) - 1}{n_i}  ,
\]
where the sum is over partitions $\fp$ with $N$ total repetitions,
and with each $p_i \in R_3$.
\end{theorem}

Like theorem \ref{thm:master}, this theorem also gives a method
for computing something which would otherwise have seemed elusive.
In this case we can plot, for any particular dimension $d$, the
fraction $f_d(N)$ of all representations in dimension $d$ which
preserve an $N$-dimensional noiseless subsystem, as a function of
$N$. The results (figs.~\ref{Fig:ss76} and \ref{Fig:ss100}) are
illuminating.

First note from the definition of $f_d(n)$ that
\[
    \sum_{n=1}^d f_d(n) = 1,
\]
since every representation is counted once. Therefore, we may view
$f_d(n)$ as the probability distribution function for a random
variable. Further, if $d \gg 1$, then the collection of points $\{
(n, f_d(n)) : n=1\ldots d\}$ forms a smooth curve,\footnote{For
$\su_3$, it is sufficient to take $d > 50$ to see the smooth
shape.} which must then subtend unit total area; see figure
\ref{Fig:ss76}.

Figure \ref{Fig:ss100} shows the exact values of $f_{100}(n)$
together with a fit to the standard \emph{inverted beta} statistical
distribution, defined by
\bel{eq:ibeta}
    f(x) = {n_{\alpha,\beta}}^{-1}\, x^{\alpha-1} (1+x)^{-\alpha-\beta}
\ee
where $\alpha,\beta>0$ and $n_{\alpha,\beta} = \int_0^1 x^{\alpha-1}
(1-x)^{\beta-1} dx$ is a normalization constant.

Since the distribution $f_d(n)$ is not mathematically known to take
the form \er{eq:ibeta}, it is of interest to determine how close the
agreement is. It is important that the points in Figure
\ref{Fig:ss100} are not data; they are exact values for
\emph{fractions}. Therefore, the relevant ``scale'' in the plot is
the dimensionless number 1, and a particularly meaningful measure of
accuracy is simply $\Delta_f := (1/d) \sum_{i=1}^d |y_i - f(x_i)|$,
the average deviation of the model from the ``data,'' measured with
respect to the only meaningful ``scale,'' which is 1. For the
example depicted in Figure \ref{Fig:ss100}, we compute $\Delta_f =
5.2 \times 10^{-4}$.

This agreement is remarkable, and seems to improve as $d$ increases;
therefore, we come to the unexpected conclusion that $f_d(n)$ is
asymptotically (for $d \gg 1$) equivalent to an inverted beta
distribution!

Further, inspection of many examples shows that for sufficiently
large $d$, the shape of the function $f_d(n)$ does not depend
strongly on $d$. More precisely, as $d$ is increased, the function
receives overall scale factors for the horizontal and vertical axes.
In all observed examples, the curve always has a peak at around
$d/5$. Given that the distribution function $f_d(n)$ is completely
insensitive to the details of the system Hamiltonian $H_S$ and to
the details of the heat bath to which the system is coupled, and
displays a simple scaling behavior with respect to the dimension of
the system Hilbert space, one is tempted to term this ``universal
behavior.''

\section{Conclusions and Outlook}

As soon has one has appreciated the importance of representations of
the  $\su_n$ Lie algebras in physics, natural questions arise. How
many of these representations exist, up to equivalence, for fixed $d
= \dim(\cH)$? How does the number grow with $d$? Suppose that we
reduce each representation into irreducibles. What is the
statistical distribution function which describes the relative
frequencies of the different possible reductions? The present work
has given answers (in some cases only partial) to each of these
questions, while interpreting the results within the unified
understanding of quantum decoherence that has emerged in the last
decade \cite{Kempe:00,ritter:05}. Many unanswered questions remain,
such as a proof of Conjecture \ref{conj}, a better explanation of
Fig.~\ref{Fig:Convergence} than the heuristic one given in
Sec.~\ref{sec:singlets}, analogues of these results for other Lie
algebras, and a fundamental derivation of the ``$d/5$ rule,'' by
which we mean the curious fact that the curves plotted in
Sec.~\ref{sec:subsystems} seem to always peak around $N = d/5$.

\subsection*{Acknowledgements}

The author gratefully acknowledges helpful discussions with Gregg
Zuckerman and Noam Elkies, and helpful comments by the referee on
the first draft.

\appendix

\section{Computing the number of irreducible representations in any dimension} \label{AppA}

In this appendix, we describe how to efficiently compute the
function $\xi(n)$ that was used elsewhere in the paper.

For a Young diagram $\lambda$, we let $d ( \lambda,\mathfrak{g})$
denote the dimension of the corresponding irreducible module of
the Lie algebra $\mathfrak{g}$. Unless noted otherwise, we will
henceforth assume $\mathfrak{g}= \tmop{su}_r$, and denote the
dimension by $d ( \lambda, r )$. Further, we often set $r = 3$, in
which case we will suppress $r$ from the notation. Young diagrams
for $\tmop{su}_3$ are characterized by row lengths $n_1$, $n_2$
with $n_1 \geq 0$ and $0 \leq n_2 \leq n_1$. The fundamental
representation is $\lambda = ( 1, 0 )$ while the adjoint is
$\lambda = ( 2, 1 )$. The representations $( n_1, n_2 )$ and $(
n_1, n_1 - n_2 )$ are conjugate to each other and have the same
dimension. Representations of the form $( 2 n, n )$ are
self-conjugate.

The dimension given by the Weyl character formula
\cite{HarrisFulton} is then
\begin{align*}
  d ( n_1, n_2 ) &= \frac{1}{2} ( n_1 + 2 ) ( n_2 + 1 ) ( n_1 - n_2 + 1 )
  \cr
  &=
   x y ( x - y )/2
\end{align*}
where $x = n_1 + 2$, and $y = n_2 + 1$.

We now compute the total number $N_D$ of irreducible
representations of $\su_3$ with dimension less than $D$, by
finding, for each fixed $n_1$, the number of $n_2$ which satisfy
$d ( n_1, n_2 ) \leq D$, and then sum over $n_1$. Expressing the
sum in terms of $x = n_1 + 2$ gives simpler notation. We give the
result as theorem \ref{thm:ND}, omitting the lengthy but
straightforward proof. For a real number $\gamma \in \R$, we let
$\lfloor \gamma \rfloor$ denote the greatest integer less than
$\gamma$. Similarly, $\lceil \gamma \rceil$ denotes the least
integer above $\gamma$.

\begin{theorem} \label{thm:ND}
Let $N_D$ denote the total number of irreducible representations
of $\su(3)$ with dimension less than $D$. Then $N_D$ is given
\emph{exactly} by
  \bel{nd}
    N_D
      = \frac{1}{2} l_D ( l_D - 1 ) - \sum_{x = k_D + 1}^{l_D} (
      \lfloor y_+ \rfloor - \lceil y_- \rceil + 1 ) .
   \ee
where $k_D : = \lfloor 2 \sqrt[3]{D} \rfloor$, $l_D := \lfloor
\frac{1}{2} ( 1 + \sqrt{1 + 8 D} ) \rfloor$,  and
\[
    y_{\pm} =
    \f{x - 2}{2} \pm \frac{1}{2} \lrp{ x^2 - 8 D / x }^{1 / 2} .
\]
\end{theorem}

\begin{corollary}
The exact number $\xi ( D )$ of irreducible representations in
dimension $D$ is given by
\begin{equation} \label{eqn:xi}
  \xi ( D ) = N_{D + \frac{1}{2}} - N_{D - \frac{1}{2}} .
\end{equation}
\end{corollary}

For determining the number of irreducible representations of
dimension $D$, eqns.~\er{nd}-\er{eqn:xi} provide a radical
computational speedup over the naive algorithm of enumerating all
possible Young diagrams and computing the dimension of each. These
equations may be implemented with an optimized C program.

\bibliographystyle{apsrev}

\end{document}